\documentclass[a4paper,twocolumn,amsmath]{revtex4-1}
\usepackage{soul}

\usepackage{adjustbox}
\usepackage{color}
\usepackage{graphicx}
\usepackage{subfigure}
\usepackage{amsmath}
\usepackage{amsfonts}
\usepackage{citesort}
\usepackage{bm}

\newcommand{\eg}{{e.\,g.\ }}
\newcommand{\ie}{{i.\,e.\ }}

\newcommand{\HH}{\mathcal{H}}
\newcommand{\UU}{\mathcal{U}}
\newcommand{\TT}{\mathcal{T}}
\newcommand{\VV}{\mathcal{V}}

\newcommand{\imag}{\mathrm{i}}
\newcommand{\total}{\mathrm{d}}

\newcommand{\multiint}[2]{{\int_0^{#1} \!\!\! \cdots \total {#1}_{\{#2\}}}}

\begin{document}

\author{ C.~Wolff }
\email{cwo@mci.sdu.dk}
\affiliation{Center for Nano Optics, University of Southern Denmark,
Campusvej 55, DK-5230 Odense M, Denmark}

\author{ C.~Tserkezis }
\affiliation{Center for Nano Optics, University of Southern Denmark,
Campusvej 55, DK-5230 Odense M, Denmark}

\author{ N.~A. Mortensen }
\affiliation{Center for Nano Optics, University of Southern Denmark,
Campusvej 55, DK-5230 Odense M, Denmark}
\affiliation{Danish Institute for Advanced Study, University of Southern
Denmark, Campusvej 55, DK-5230 Odense M, Denmark}

\title{On the time evolution at a fluctuating exceptional point}

\date{\today}
  
\begin{abstract}
  We theoretically evaluate the impact of drift-free noise on the dynamics of
  $\mathcal{PT}$-symmetric non-Hermitian systems with an exceptional point, which
  have recently been proposed for sensors. 
  Such systems are currently considered as promising templates for sensing applications,
  because of their intrinsically extremely sensitive response to external perturbations.
  However, this applies equally to the impact of fabrication imperfections and
  fluctuations in the system parameters. 
  Here we focus on the influence of such fluctuations caused by inevitable (thermal) noise
  and show that the exceptional-point eigenstate is not stable in its presence. 
  To this end, we derive an effective differential equation for the mean time evolution
  operator averaged over all realizations of the noise field and via numerical analysis
  we find that the presence of noise leads to exponential divergence of any initial state 
  after some characteristic period of time.
  We therefore show that it is rather demanding to design sensor systems based on
  continuous operation at an exceptional point.
\end{abstract}

\maketitle

\section{Introduction}\label{Sec:Intro}

In the recent study of
parity-time ($\mathcal{PT}$)-symmetric non-Hermitian dynamic 
systems~\cite{Bender:1998,El-Ganainy:2018,Longhi:2017}, the notion of 
exceptional points~\cite{Guo:2009,Rueter:2010,Regensburger:2012,Brandstetter:2014,Peng:2014,Hodaei:2014} has attracted 
particular interest, \eg for the realization of highly sensitive 
sensors~\cite{Miller:2017}.
A typical realization of such a system would consist of coupled oscillators 
(\eg evanescently coupled optical resonators) where one oscillator is subject to
gain and the other to an equal amount loss~\cite{Hodaei:2017}.
An exceptional point in the space formed by the parameters ``gain'' and 
``coupling strength'' is characterized by the fact that not only two (or even 
more) eigenvalues are degenerate, but that also their eigen\emph{states} 
coalesce.
In this sense it bears great similarity to the critically damped harmonic 
oscillator, which is the optimal operating point for various types of sensing 
equipment such as galvanometers~\cite{Bajaj:2004}.
Similarly, a coupled-oscillator system operated at an exceptional point is in itself 
particularly well-suited for sensing applications, because any small 
perturbation $\Delta$ leads to a splitting of the eigenvalues that scales with 
the square root of $\Delta$~\cite{Wiersig:2014,Hodaei:2017}.
In other words, exceptional points promise the design of extremely sensitive
sensor configurations.

However, this extreme sensitivity is as much of a curse for practical purposes
as it is a blessing, because even minuscule deviations of the operating point
move the system away from the exceptional point and thus diminish the high
sensitivity.
It is clear that both imperfections during manufacturing as well as drift of 
the operating point are a major concern and that they must be compensated by
the introduction of an active stabilization of the operation point via a
feedback amplifier similar to chopper-stabilized operational amplifiers.
Ideally, this would involve to operate the sensor in its stationary state at 
the exceptional point.
However, since the time evolution of such systems includes a linearly growing
contribution~\cite{Wiersig:2008,Heiss:2010,Heiss:2012}, 
it is not clear from the outset how the system would 
react to inevitable noise.
In this context, we define drift as any fluctuation that is eliminated by a 
stabilization circuit and refer to the remaining fluctuations as noise in a 
strict sense.
It should be stressed that by this definition, the noise spectrum has a low-frequency 
gap around the value $\omega=0$, which will become crucial for the following.
Furthermore, we remark that any stabilization circuit will itself contribute 
noise.

The problem of imperfections and noise on the performance of hypothetical exceptional-point
sensors has been addressed before in recent literature.
One major contribution to this was the realization that while systems at an
exceptional point do exhibit a square-root law for the eigenvalue splitting in response
to small perturbations, this does not lead to an improvement in the ratio between
the signal and the fundamental quantum noise level~\cite{Langbein:2018}.
In other words, it has been shown that an exceptional point does not provide a benefit
in the quantum-noise limited regime, although it might make it easier to reach this
limit.
A different angle was approached by our group in a recent paper on the effect
of sample-to-sample variations~\cite{Mortensen:2018}, where we discuss (among other
things) how drift in the system parameters leads to an exponentially growing error 
in the state of an exceptional point sensor.
In this paper, we go beyond this preexisting work and show that eliminating drift does
not eliminate this divergent behavior. 
We show that drift-free fluctuations in the system parameters (e.g. the site detuning
due to inevitable thermal fluctuations of the resonator geometry or fluctuations in 
the gain) around the exceptional point lead to an exponential divergence of the state
error.
As a consequence, it is impossible to operate any real-world system at the
exceptional point for a large period of time.

The paper is structured as follows:
In Section 2, we introduce the problem and define our notation and
conventions.
In Section 3, we present the ordinary differential equation that
describes the time evolution of the noisy system averaged over all realizations
of the noise field.
In Section 4, we present a brief summary of the analytical derivation that
leads to this differential equation and compare its solutions to brute-force 
numerical calculations for some particular realizations of the noise spectrum.
In Section 5, we discuss the consequences of our findings for the design and
feasibility of optical sensors based on exceptional-point dynamics,
and discuss the prospects of exceptional point-based sensing in view of our analysis.
After the Conclusion,
the paper ends with two appendices
with additional details about the solution and its numerical implementation.

\section{Preliminaries}

We study the time evolution of a two-site $\mathcal{PT}$-symmetric system at an 
exceptional point.
Within a coupled-mode picture, the dynamics of any such system is described by 
the equation
\begin{align}
  \imag \partial_t \begin{pmatrix} a_1(t) \\ a_2(t) \end{pmatrix} 
  = \begin{pmatrix} \omega - \imag g & \kappa \\ 
  \kappa & \omega + \imag g \end{pmatrix}
  \begin{pmatrix} a_1(t) \\ a_2(t) \end{pmatrix},
\end{align}
where $a_1(t)$ and $a_2(t)$ are the complex amplitudes of the 
respective resonator modes, $\omega$ is their common eigenfrequency, $g$ 
is the gain or loss (depending on the sign) that they are subjected to and $\kappa$ is
the coupling  constant, which was chosen to be real by an appropriate choice for the
relative phase between the modes.
Next, we switch to a frame of reference that rotates with the phase 
$\exp(-\imag \omega)$ and introduce a rescaled time variable 
$\tau = \kappa t$.
The latter means that we measure time in units of the inverse coupling 
constant.
Thus, we then find the equation of motion
\begin{align}
  \imag \partial_\tau \psi(\tau) = \begin{pmatrix} -\imag g / \kappa & 1 \\ 
  1 & \imag g / \kappa \end{pmatrix} \psi(\tau),
\end{align}
where the state vector $\psi(\tau) = 
\exp(-\imag \omega t / \kappa) [a_1(t), a_2(t)]^T$ 
comprises the mode amplitudes in the rotating frame.
This system has an exceptional point for $g = \kappa$.
The exceptional-point dynamics of every two-site $\mathcal{PT}$-symmetric system 
can be thus reduced to the prototypical Hamiltonian 
\begin{align}
  \HH_0 = & 
  \begin{pmatrix} -\imag & 1 \\ 1 & \imag \end{pmatrix}
    =
  \sigma_x - \imag \sigma_z,
\end{align}
where $\sigma_i$ denote the Pauli matrices and the time evolution of a state 
$\psi(\tau)$ of this ideal system is given by a Schr\"odinger-type equation
\begin{align}
  \partial_\tau \psi(\tau) = - \imag \HH \psi(\tau),
\end{align}
with respect to the rescaled time variable $\tau$ of the transformed system.

We now assume that the operating point of the system is perturbed by some 
time dependent real-valued fluctuation $\Delta(\tau)$, which can be represented 
as a Fourier integral
\begin{align}
  \Delta(\tau) = \int_{-\infty}^\infty \total \omega \ 
  b(\omega) \exp(-\imag\omega\tau).
  \label{eqn:noise_def}
\end{align}
The phase of the function $b(\omega)$ is assumed to fluctuate randomly and 
arbitrarily quickly in $\omega$ while its modulus is a smooth function of 
$\omega$~\cite{BornWolf:1999}.
It is connected to the fluctuation power spectrum $P(\omega)$ (again in 
appropriately chosen dimensionless units):
\begin{align}
  P(\omega) = |b(\omega)|^2.
\end{align}
We assume the overall fluctuation power to be finite, so $P(\omega)$ must 
cut off at high frequencies.
Furthermore, we distinguish between low-frequency and quasi-static 
fluctuations, which we call drift, and high-frequency fluctuations, which
we call noise.
The former are assumed to be eliminated by an active stabilization circuit,
with only the latter remaining.
In other words, we assume that the relevant fluctuation field $\Delta(\tau)$
vanishes in a neighborhood of $\omega=0$, if only to prevent the system from 
permanently drifting away from the exceptional point.
We assume
\begin{align}
  P(\omega) = 0 \quad \text{for} \quad |\omega| < \omega_\text{min}.
\end{align}
The key characteristic of the noise function within this paper is the
auto-correlation function $\Gamma(\tau)$.
Since the auto-correlation of a white noise is a sharp peak at $\tau=0$, we 
approximate it as a Dirac distribution:
\begin{align}
  \Gamma(\tau) = \int_{-\infty}^\infty \total \tau' 
  \Delta(\tau') \Delta(\tau' + \tau) \approx \gamma \delta(\tau),
  \label{eqn:autocorrelation}
\end{align}
with some constant $\gamma$  formally given by $\gamma=\iint_{-\infty}^\infty \total \tau' \total \tau
  \Delta(\tau') \Delta(\tau' + \tau)$.

In the following, we assume that the fluctuation field detunes the on-site 
energies of the two coupled sites, \ie we introduce a perturbation operator
$\VV(\tau) = \Delta(\tau) \sigma_z$.
A fluctuation in the gain and loss parameters of the two-site problem can be 
described by a second operator 
$\VV'(\tau) = \imag \Delta'(\tau) \sigma_z$ generated by a second 
fluctuation field $\Delta'(\tau)$.
This leads to results that differ from the ones obtained for $\VV(\tau)$
by only the imaginary unit and [assuming no correlations between $\Delta(\tau)$
and $\Delta'(\tau)$] their respective corrections to the total time evolution 
can be simply added.
Therefore, it is sufficient to study the problem of fluctuating on-site energy 
detuning:
\begin{align}
  \HH(\tau) = & \HH_0 + \Delta(\tau) \sigma_z.
\end{align}

This type of problem requires heavy use of nested time integrals for which we
introduce a short-hand notation:
\begin{align}
  \nonumber
  \multiint{\tau}{n}
  \ = \
  \int_0^{\tau} \total \tau_1 \int_0^{\tau_1} \total \tau_2
  \cdots \int_0^{\tau_{n-1}} \total \tau_n .
\end{align}

%%%%%%%%%%%%%%%%%%%%%%%%%%%%%%%%%%%%%%%%%%%%%%%%%%%%%%%%%%%%%%%%%%%%%%%%%%%%%%%%
\section{Main result}

The topic of this study is the derivation of the mean time evolution of an 
exceptional-point system in the presence of a noisy perturbation to the 
system parameters, averaged over all possible realizations of the noise field.
The natural description for the dynamics of such a system is the time evolution
operator $\bar\UU(\tau)$, which we find to satisfy the ordinary initial value problem
\begin{subequations}
\label{eq:mainresult}
\begin{align}
  \bar\UU''(\tau) = & -\gamma [1 + \imag \HH_0^\dagger \tau] 
  \bar\UU'(\tau) + 2 \gamma [\sigma_z - \tau] \bar\UU(\tau),
  \label{eqn:avg_ode1}
  \\
  \bar\UU'(0) = & -\imag\HH_0,
  \label{eqn:avg_ode2}
  \\
  \bar\UU(0) = & 1,
  \label{eqn:avg_ode3}
\end{align}
\end{subequations}
where $\tau$ is the dimensionless time variable, $\HH_0^\dagger$ is the adjoint  of the
(non-Hermitian) Hamiltonian at the unperturbed exceptional point and $\gamma$ is the
parameter of the noise autocorrelation function as introduced in
Eq.~\eqref{eqn:autocorrelation}.

The solution to Eqs.~(\ref{eqn:avg_ode1}--\ref{eqn:avg_ode3}) at first follows the
noiseless dynamics, which means that the norm of the exceptional-point eigenstate
remains constant and the norm of non-eigenstates grows linearly.
After some characteristic time the system enters a new regime, where the norm of any
initial state grows exponentially, which can be expressed to very good accuracy by
the equation
\begin{align}\label{eq:approx}
  |\Delta\!\psi(\tau)| \approx \exp\left( \sqrt{2 \gamma} \tau - 1 \right).
\end{align}

This means that a system with noisy system parameters can be operated at the
exceptional point for a time no longer than $\tau_\text{max} = 1/\sqrt{2\gamma}$.
At this point, any active stabilization circuit will kick in and try to stabilize the
norm of the state, moving one system parameter away from the exceptional point.
Depending on the characteristics of the feedback circuit, the system will either
settle at this new equilibrium point or the feedback circuit will become unstable and
enter oscillations.
The growing sensitivity close to the exceptional point suggests that a reduction in
noise (and hence in distance between the equilibrium operating point and the exceptional
point) comes at the price of increased tendency for oscillation of the system parameters
(detuning, gain or loss).

%%%%%%%%%%%%%%%%%%%%%%%%%%%%%%%%%%%%%%%%%%%%%%%%%%%%%%%%%%%%%%%%%%%%%%%%%%%%%%%%
\section{Sketch of the derivation}

The time evolution of a state including a fluctuating perturbation of the
exceptional point is described by the equation
\begin{align}
  \partial_\tau \psi(\tau) = & -\imag \HH(\tau) \psi(\tau) 
  = -\imag [\HH_0 + \Delta(\tau) \sigma_z] \psi(\tau),
  \label{eqn:full_schroedinger_eqn}
\end{align}
which,
in analogy to the treatment within the Heisenberg representation in
quantum mechanics~\cite{Sakurai:1994,Baym:1990}
is conventionally solved by the time evolution operator given as a 
Neumann series
\begin{align}
  \UU(\tau) = & 1 + \sum_{n=1}^\infty \TT^{(n)}(\tau),
  \label{eqn:neumann_series_orig}
\end{align}
where the summand $\TT^{(n)}(\tau)$ are the orders in the time variable $\tau$:
\begin{align}
  \TT^{(n)}(\tau) 
  = & \multiint{\tau}{n} \ 
  \prod_{j=1}^n (-\imag)[\HH_0 + \Delta(\tau_j) \sigma_z].
\end{align}
With this, the time evolution operator satisfies the 
Dyson series-like~\cite{Sakurai:1994}
self-consistent equation
\begin{align}
  \nonumber
  \UU(\tau) = & 1 - \imag \HH_0 \tau 
  - \imag \multiint{\tau}{2} \ 
  [\delta(\tau_1 - \tau_2) - \imag \HH_0] 
  \\
  & \quad \times
  \sigma_z \Delta(\tau_2) \UU(\tau_2).
\end{align}
This expansion is ill-suited for our goal to derive the mean time evolution 
operator $\bar\UU(\tau)$ averaged over all realizations of the noise function 
$\Delta(\tau)$.
This is because it does not separate the odd powers of the perturbation
$\VV(\tau)$, which all average to zero, from the even powers, whose averages
have finite values (see Appendix \ref{appx:noise_properties}).
Therefore, we reorder the series to find the equivalent equation
\begin{align}
  \nonumber
  \UU(\tau) 
  = & 1 - \imag \HH_0 \tau 
  - \imag \multiint{\tau}{2} \ 
  [\delta(\tau_1 - \tau_2) - \imag \HH_0] 
  \\
  \nonumber
  & \quad \times
  \sigma_z \Delta(\tau_2) 
  [1 - \imag \HH_0 \tau_2]
  \\
  & \quad 
  \nonumber
  - \multiint{\tau}{4} \ 
  [\delta(\tau_1 - \tau_2) - \imag \HH_0] \sigma_z \Delta(\tau_2) 
  \\
  & \quad \times 
  [\delta(\tau_3 - \tau_4) - \imag \HH_0] \sigma_z \Delta(\tau_4) 
  \UU(\tau_4).
\end{align}
The solution to this equation depends critically on the precise function 
$\Delta(\tau)$, which is of course unknown.
Instead, we only know statistical properties such as the power spectrum 
moments and the autocorrelation function [Eq.~\eqref{eqn:autocorrelation}].
Therefore, the natural quantity to investigate is the average $\bar\UU(\tau) $ 
of the evolution operator $\UU(\tau)$ over all realizations of $\Delta(\tau)$.
Next, we assume that the different realizations of $\Delta(\tau)$ are in fact
just an explicit dependence on the absolute start point $\tau_0$ of the time
evolution.
In other words, a change of the realization of the noise function 
$\Delta(\tau)$ is equivalent to a shift of the temporal origin 
$\Delta(\tau) \rightarrow \Delta(\tau+\tau_0)$.
\begin{widetext}
The evolution operator associated with start time $\tau_0$ satisfies:
\begin{align}
  \nonumber
  & \UU_{\tau_0}(\tau) 
  =  1 - \imag \HH_0 \tau 
  - \overbrace{
      \imag \multiint{\tau}{2} \ 
    [\delta(\tau_1 - \tau_2) - \imag \HH_0] 
    \sigma_z \Delta(\tau_2 - \tau_0) 
    [1 - \imag \HH_0 \tau_2]
  }^{\text{averages to zero when integrated over $\tau_0$}}
  \\
  & 
  - \multiint{\tau}{4} \ 
  [\delta(\tau_1 - \tau_2) - \imag \HH_0] \sigma_z \Delta(\tau_2 - \tau_0) 
  [\delta(\tau_3 - \tau_4) - \imag \HH_0] \sigma_z \Delta(\tau_4 - \tau_0) 
  \UU_{\tau_0}(\tau_4),
  \label{eqn:time_evol_realization_recursion}
\end{align}
\end{widetext}
and the realization-averaged evolution operator $\bar\UU(\tau)$ 
is given as an integral over $\tau_0$:
\begin{align}
  \bar\UU(\tau) = \int_{-\infty}^\infty \total \tau_0 \ 
  \UU_{\tau_0}(\tau).
  \label{eqn:avg_time_evol_def}
\end{align}
To interpret the realization dependence as an explicit time dependence
may seem a bit unconventional at first, but it is just the reverse of the 
usual strategy in statistical physics to replace time averages with ensemble 
averages assuming quasi-ergodicity.
Assuming that the fluctuation function $\Delta(\tau)$ is the result of a 
(\eg thermodynamic) process that satisfies quasi-ergodicity justifies our 
choice.

Next, we substitute Eq.~\eqref{eqn:avg_time_evol_def} in 
Eq.~\eqref{eqn:time_evol_realization_recursion} and replace 
$\UU_{\tau_0}(\tau)$ with $\bar\UU(\tau)$ under the integral.
This essentially constitutes a random-phase approximation and should be a 
decent approximation if the noise spectrum is negligible in the frequency range 
of the coupling parameter (\ie frequencies comparable with the Rabi frequency 
away from the exceptional point).
Finally, we drop the first integral from the recursion equations as annotated
in Eq.~\eqref{eqn:time_evol_realization_recursion}.
This assumption holds for evolution times $\tau$ that are large compared to
the lower cut-off frequency of the noise spectrum, because the moments of
$\langle \tau^n \Delta(\tau)\rangle = 0$ 
(see Appendix~\ref{appx:noise_properties}).
We find for the realization-averaged evolution operator:
\begin{align}
  \nonumber
  \bar\UU(\tau) \approx & 
%  1 - \imag \HH_0 \tau
%  \\ 
%  & \quad \nonumber
%  - \int_{-\infty}^\infty \total \tau_0 \ \multiint{\tau}{4} \ 
%    [\delta(\tau_1 - \tau_2) - \imag \HH_0] 
%  \\ 
%  & \quad \times \nonumber
%    \sigma_z \Delta(\tau_2 - \tau_0) [\delta(\tau_3 - \tau_4) - \imag \HH_0] 
%  \\ 
%  & \quad \times
%  \sigma_z \Delta(\tau_4 - \tau_0) 
%  \bar \UU(\tau_4)
%  \\
%  \nonumber
%  = & 
  1 - \imag \HH_0 \tau
  - \multiint{\tau}{4} \ 
    [\delta(\tau_1 - \tau_2) - \imag \HH_0] \sigma_z  
  \\
  & \quad \times 
    [\delta(\tau_3 - \tau_4) - \imag \HH_0] 
  \sigma_z \Gamma(\tau_2 - \tau_4) 
  \bar \UU(\tau_4),
  \label{eqn:avg_int_Gamma}
\end{align}
where $\Gamma(\tau)$ is the auto-correlation function as introduced in 
Eq.~\eqref{eqn:autocorrelation}.
\begin{figure}
  \includegraphics[width=1.0\columnwidth]{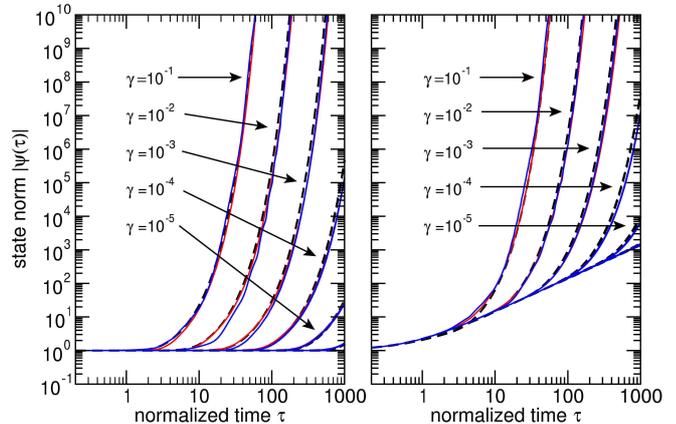}
  \caption{
    Left panel:
    Double-logarithmic plot of the norm $|\psi(\tau)|$ of the state starting 
    with the exceptional-point eigenstate $\psi(0) = (1, \imag)^T / \sqrt{2}$
    for different values of $\gamma$ (see graph annotations).
    The black dashed curves are the numerical solutions to the effective
    differential equation~(\ref{eqn:avg_ode1}--\ref{eqn:avg_ode3}) using 
    fourth-order Runge--Kutta integration.
    The red and blue solid lines are brute-force calculations of 
    Eq.~\eqref{eqn:full_schroedinger_eqn} modelling the noise as two 
    quite different ensembles consisting of 1000 harmonic oscillators each
    (red curve: high-frequency noise; blue curve: low-frequency noise; 
    see Appendix~\ref{appx:numerics} for details). \\
    Right panel: Same as left panel starting with the non-eigenstate 
    $\psi(0) = (0, 1)^T$.
  }
  \label{fig:numerical_solution}
\end{figure}
We can now simplify Eq.~\eqref{eqn:avg_int_Gamma}:
\begin{align}
  \nonumber
  \bar\UU(\tau)
  = & 1 - \imag \HH_0 \tau
  - \gamma \int_0^\tau \total \tau_1 \sigma_z [1 - \imag \HH_0 \tau_1] \sigma_z
  \bar \UU(\tau_1)
  \\
  & \quad
  + \imag \gamma \multiint{\tau}{2} \HH_0 \sigma_z [1 - \imag \HH_0 \tau_2] 
  \sigma_z \UU(\tau_2).
  \label{eqn:avg_time_evol_recursion}
\end{align}
This can be transformed to the ordinary initial value problem that we have already stated in Eq.~(\ref{eq:mainresult}).

%%%%%%%%%%%%%%%%%%%%%%%%%%%%%%%%%%%%%%%%%%%%%%%%%%%%%%%%%%%%%%%%%%%%%%%%%%%%%%%%
\section{Results and Discussion}
The full analytical solution for equations like Eq.~(\ref{eq:mainresult}) with scalar
coefficients is a product between the exponential and the hypergeometric function.
A matrix generalization of this with non-commuting arguments would be beyond the 
scope of this paper.
Nonetheless, a numerical solution is straightforward and compared in 
Fig.~\ref{fig:numerical_solution} to brute-force solutions of the full problem 
for two quite different realizations of the noise function $\Delta(\tau)$.
Further details on the numerics employed can be found in 
Appendix~\ref{appx:numerics}.

\begin{figure}
  \includegraphics[width=1.0\columnwidth]{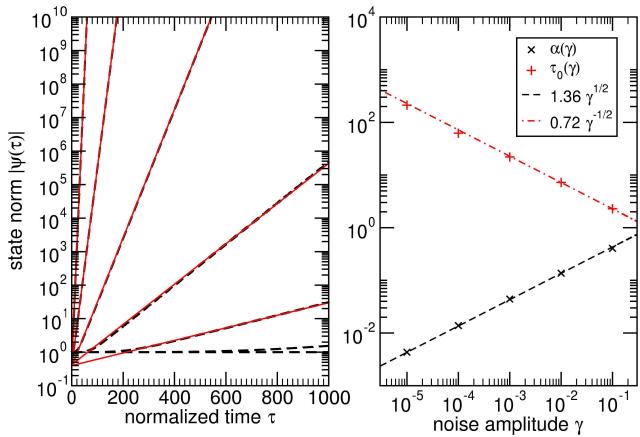}
  \caption{
    Left panel:
    Time-evolution of the effective initial value problem
    [Eqs.~(\ref{eqn:avg_ode1}--\ref{eqn:avg_ode3}), dashed black lines for 
    values of $\gamma$ annotated in Fig.~\ref{fig:numerical_solution}] starting
    from the exceptional-point eigenstate $\psi(0) = (1, \imag)^T/\sqrt{2}$ 
    compared to expressions of the form $f(\tau) = \exp[\alpha(\tau - \tau_0)]$ 
    (red solid lines) with parameters $\alpha(\gamma)$ and $\tau_0(\gamma)$ 
    determined by least-squares fitting.
    Clearly, the long-term behavior is an exponential growth; the parameter 
    $\tau_0$ is a measure for the time over which the time evolution mostly
    follows the unperturbed dynamics.\\
    Right panel: 
    Behavior of the functions $\alpha(\gamma)$ and $\tau_0(\gamma)$ as 
    extracted from the left panel graphs.
    Clearly, both quantities are proportional to $\gamma^{1/2}$ and 
    $\gamma^{-1/2}$, respectively, over a large dynamic range.
  }
  \label{fig:scaling}
\end{figure}

Both the fully numerical examples and our effective description show that the
time evolution of an initial state at first follows the behavior expected for
the noiseless Hamiltonian.
This is a stationary evolution for the eigenstate 
$\psi(0) = (1, \imag)^T / \sqrt{2}$ and linearly growing norm for
any non-eigenstate, \eg $\psi(0) = (0, 1)^T$.
After a characteristic time $\tau_0$ that depends on the noise amplitude 
$\gamma$, the system enters an exponentially divergent regime.
This is more clearly seen in the semi-logarithmic plot in the left-hand panel
of Fig.~\ref{fig:scaling}.
Here, we compare the effective numerical solution of the time evolution of the
exceptional-point eigenstate to the simple Ansatz
\begin{align}
  f(\tau) = \exp[\alpha (\tau - \tau_0)],
\end{align}
where $\alpha$ describes how quickly the state diverges 
from
the expected
behavior and $\tau_0$ after which time the divergent regime sets in.
As we show in the right-hand panel of Fig.~\ref{fig:scaling}, the 
$\gamma$-dependence of both quantities is reasonably well described by simple
square-root laws: $ \alpha(\gamma) =  \sqrt{2 \gamma}$ and $\alpha \tau_0(\gamma) =  1$.
This is the result summarized in Eq.~(\ref{eq:approx}).

One might wonder what values for the characteristic time can be expected 
in an actual experiment.
This is fairly hard to answer without knowledge of the possible
origins of noise and the respective amplitudes in a given experiment.
However, we are able to make some very rough estimates based on the inevitable
noise of the pump laser in optically pumped $\mathcal{PT}$-symmetric microring
dimers with a few nanometer resonance splitting of the symmetrically pumped resonators such as those presented in Ref.~\cite{Hodaei:2014}.
The natural time unit is the inverse coupling parameter $\kappa$, which in this example is of the order $\kappa \approx 10^{12}\,\text{s}^{-1}$, \ie $\tau = 1$ is on the
order of picoseconds.
The fluctuations of the gain parameter relative to the mean gain
can be roughly estimated as identical to the relative intensity fluctuations of the pump.
According to the well-known Wiener--Khinchine theorem~\cite{BornWolf:1999}, this number is given by the relative spectral power density of the wide band intensity fluctuations.
A realistic intensity noise figure for a small lasers is around $-120\,\text{dBc}$.
Therefore, we find the rough order of magnitude $\gamma \approx 10^{-12}$,
which is well within the range of validity for our perturbative expansion; we also
would like to point out that this range of $\gamma$ comes close to double precision
machine accuracy and that brute-force simulations of Eq.~\eqref{eqn:full_schroedinger_eqn} would thus not be trustworthy for this value.
From Fig.~\ref{fig:scaling}, we can then estimate $\tau \approx 10^6$, which
in real time corresponds to the order of microseconds.
By reducing the noise figure, this can be of course increased with a square root law.

In view of our present analysis, one would naturally inquire if it is possible
to design systems that fully take advantage of the fine sensitivity of exceptional
points while at the same time eliminating the instabilities discussed here.
While we cannot at the moment foresee a practical solution, topologically protected
structures with exceptional points have been attracting increasing interest in recent
years~\cite{Shen:2018,Gangaraj:2018}, and in \cite{Gangaraj:2018}, in particular,
it was shown that they can be very efficient in dealing with imperfections and loss in
the case of waveguiding. It is not unreasonable therefore to speculate that they could
potentially offer a route towards a solution in the case of sensing as well.

\section{Conclusions}
We have provided a detailed analytical study of the dynamics of a 
$\mathcal{PT}$-symmetric two-site coupled-mode system at the exceptional point, subject
to drift-free fluctuations in its system parameters.
To this end, we have analytically derived an effective differential equation that 
describes the mean time-evolution operator of this type of system.
The fluctuations are assumed to be due to (\eg thermal) noise, where the quasi-static
contributions (drift) have been eliminated by means of an external stabilization 
system.
The numerical solution of the effective differential equation shows that the presence of
noise leads inevitably to the exponential divergence both of the noiseless system's 
eigenstate as well as non-eigenstates.
As we find, the divergence occurs on a time scale that depends on the noise amplitude.
The numerical solutions of the effective model are in excellent agreement with
brute-force simulations that we performed by modelling the noise as the result of a 
bath of incoherent harmonic oscillators.
This implies that harnessing the characteristic dynamics at an exceptional point for
the design of highly sensitive sensors in practical applications faces not only the
challenge that the quantum-noise limit cannot be overcome and that the delicate
balance of system parameters is extremely sensitive to drift, but also that 
stabilization measures to keep the system at the exceptional point are exceedingly
prone to amplifier noise and might suffer from regulation instabilities.
Maintaining operation at the exceptional point for long enough times to detect 
minute resonance splittings seems to require very careful design of the feedback
system.
We believe that our effective differential equation for the time evolution at noisy
exceptional points provides a valuable tool in this engineering feat as it can be 
extended to an analytical model for the complete system including the active 
stabilization system.
This in turn would provide insight into the underlying processes, could be analyzed
analytically \eg for overall stability and could be used to optimize a system for 
the maximally sensitive equilibrium operating point without the need for a large number
of brute-force simulations involving different realizations of the noise field.

\section{Acknowledgments}

N.~A.~M. is a VILLUM Investigator supported by VILLUM Fonden (grant No. 16498).
C.~W. acknowledges funding from a MULTIPLY fellowship under the Marie 
Sk\l{}odowska-Curie COFUND Action (grant agreement No. 713694).
C.~T. is a parasite feeding off both.
The Center for Nano Optics is financially supported by the University of 
Southern Denmark (SDU 2020 funding).

\begin{appendix}

%%%%%%%%%%%%%%%%%%%%%%%%%%%%%%%%%%%%%%%%%%%%%%%%%%%%%%%%%%%%%%%%%%%%%%%%%%%%%%%%
\section{Fundamental properties of the fluctuation field}
\label{appx:noise_properties}

We assume that the phases of $b(\omega)$ fluctuate arbitrarily 
quickly in frequency and the modulus of $b(\omega)$ decays for high $\omega$.
  As a result, all moments of $b(\omega)$ with respect to $\omega$ vanish:
\begin{align}
  \int_{-\infty}^\infty \total \omega \ \omega^n b(\omega) = 0,
  \label{eqn:noise_omega_moment}
\end{align}
for any positive exponent $n>0$.
We can generalize this to a wider class of functions:
\begin{align}
  \int_{-\infty}^\infty \total \omega \ f(\omega) b(\omega) = 0,
\end{align}
for any $f(\omega)$ that is holomorphic on the union of a finite number of 
intervals that cover the support of $b(\omega)$.
The reason is that under these conditions the integral can be decomposed into 
a finite number of integrals each covering an interval on which $f(\omega)$ 
can be represented by a Taylor series to whose terms 
Eq.~\eqref{eqn:noise_omega_moment} applies.
Thus, assuming that $b(\omega)$ vanishes in a neighborhood of $\omega=0$,
we can extend Eq.~\eqref{eqn:noise_omega_moment} to negative exponents $n$ and
also allow that the integrand be multiplied with an arbitrary entire function:
\begin{align}
  \int_{-\infty}^\infty \total \omega \ g(\omega) \omega^n b(\omega) = 0,
  \label{eqn:noise_omega_moment_allg}
\end{align}
for $g(\omega)$ entire and any integer $n \in \mathbb{Z}$.
The assumption that $b(\omega)$ vanishes around $\omega=0$ is intimately 
connected to the distinction between high-frequency fluctuations (noise) and
low-frequency fluctuations (drift). 

With this we can now show that all moments of the fluctuation field 
$\Delta(\tau)$ with respect to time vanish.
The $n$-th moment of $\Delta(\tau)$ is given as:
\begin{align}
  \langle \tau^n \Delta(\tau) \rangle
  = & \int_0^\tau \total \tau' \  \int \total \omega\, 
  b(\omega) (\tau')^n \exp(\imag\omega \tau')
  \\
  = & \int_0^\tau \total \tau' \  \int \total \omega\, 
  b(\omega)\imag^{n}\partial_{\omega}^n \exp(-\imag\omega \tau').
\end{align}
Next, we perform the temporal integral to find
\begin{align}
  \langle \tau^n \Delta(\tau) \rangle =
  \imag^{n-1} \int {\rm d}\omega\, b(\omega)\partial_{\omega}^n 
  \left[\frac{1-\exp(-\imag \omega \tau)}{\omega } \right].
\end{align}
This expression is of the type presented in 
Eq.~\eqref{eqn:noise_omega_moment_allg} and therefore vanishes:
\begin{align}
  \langle \tau^n \Delta(\tau) \rangle = 0.
\end{align}

%%%%%%%%%%%%%%%%%%%%%%%%%%%%%%%%%%%%%%%%%%%%%%%%%%%%%%%%%%%%%%%%%%%%%%%%%%%%%%%%
\section{Numerical methods}
\label{appx:numerics}

The comparison in Fig.~\ref{fig:numerical_solution} was computed numerically
in the following way.
First, the differential operator of the effective differential 
equation \eqref{eqn:avg_ode1} was brought to a first-order form
\begin{align*}
  \left[
  \partial_\tau +
  \begin{pmatrix}
    0 & 1  \\
    2 \gamma (\tau - \sigma_z) & \gamma (1 + \imag \HH_0^\dagger \tau)
  \end{pmatrix}
  \right]
\end{align*}
Then, the problem was integrated numerically using a standard 4th order 
Runge-Kutta for the two initial column vectors $(1, 0, -1, 1)^T$ and 
$(0, 1, 1, 1)^T$ equivalent to applying the conditions 
Eq.~(\ref{eqn:avg_ode2},\ref{eqn:avg_ode3}) to the physical states $(1,0)^T$
and $(0,1)^T$.
This provides the columns of $\bar\UU(\tau)$ [and as a byproduct those of 
$\bar \UU'(\tau)$]. 

This is compared to a brute-force calculation.
Random noise, being an intrinsically non-smooth signal, is not very well-suited 
for numerical integration, especially because higher-order Runge-Kutta methods
require the evaluation at different intermediate times.
Therefore, we took some inspiration from Eq.~\eqref{eqn:noise_def} and modelled
it as an ensemble of 1000 harmonic oscillators (the bath) with eigenfrequencies 
roughly equidistantly spaced in a spectral window and with random initial 
phases.
The (real-valued) amplitudes of the oscillators were added up to give a 
consistent and smooth approximation to the noise function $\Delta(\tau)$, which
was then fed into the Hamiltonian.
With this, Eq.~\eqref{eqn:full_schroedinger_eqn} was integrated in time 
alongside the ensemble of harmonic oscillators.
We show results for two frequency bands: a high frequency noise band with
bath eigenfrequencies between $3.0$ and $30.0$, \ie satisfying the assumption 
that underlies the approximation $\UU(\tau) \approx \bar\UU(\tau)$ in 
Eq.~\eqref{eqn:avg_int_Gamma}. 
The second example is for a low frequency noise band spanning from $0.3$ to
$3.0$, \ie not satisfying said assumption.
Still, our effective description seems to remain remarkably accurate.

\end{appendix}

%\bibliographystyle{...}
%\bibliography{...}
\end{document}